# Confinement Effects and Surface-Induced Charge Carriers in Bi Quantum Wires

T. E. Huber,[a] A. Nikolaeva,[b] D. Gitsu,[b] L. Konopko,[b] C. A. Foss, Jr.,[c] and M. J. Graf[d]

[a]Laser Laboratory, Howard University, Washington, DC 20059-0001;

[b]Institute of Applied Physics, Academy of Sciences, Moldova;

[c]Department of Chemistry, Georgetown University, Washington, DC; and

[d]Department of Physics, Boston College, Chestnut Hill, MA 02467

We present measurements of Shubnikov-de Haas oscillations in arrays of bismuth nanowires. For 80-nm wires, the hole concentration is less than 30% of that for bulk Bi, a finding that is consistent with current models of quantum confinement effects. However, 30-nm-diameter nanowires, which are predicted to be semiconductors, show a nearly isotropic short period of 0.025 $T^{-1}$, consistent with a heavy carrier concentration five times that of bulk Bi. These results are discussed in terms of surface-induced charge carriers in a spherical Fermi surface pocket that are uniformly distributed in the 30-nm nanowire volume and that inhibit the semimetal-to-semiconductor transition.

Bismuth is a particularly favorable material for studying the electronic properties of quantum wires due to its small electron effective mass and high carrier mobility.[1] The resonance of the Landau levels (e.g., de Haas-van Alphen [dHvA], Shubnikov-de Haas [SdH], and related effects)[2] provides an unambiguous measure of the charge density and the anisotropy of the Fermi surface (FS) of bulk crystalline Bi, which consists of three electron pockets at the L-point and a T-point hole pocket. The effective band overlap energy, $E_0$, and the Fermi energy, $E_F$, are 37 and 26 meV, respectively, levels which result in small electron and hole densities ($n_i = p_i = 3 \times 10^{17}/cm^3$ at a temperature of 2 K). Quantum confinement effects, which decrease $E_0$, become relevant for quantum wires with diameter $d \sim 2\hbar/\sqrt{2m^*E_0}$, where $m^*$ is the corresponding electron in-plane effective mass transverse to the wire axis. For wires oriented along $C_3$ (trigonal direction), $m^* = 0.0023$ and the relevant diameter is 42 nm. Detailed calculations,[1] which assume that the components of effective mass tensors do not change upon confinement, show that a semimetal-to-semiconductor (SMSC) transition occurs for $d_c \sim 55$ nm for wires oriented along the trigonal direction. Various experimental results support this theory. Recently, room-temperature infrared absorption results for 30- and 60-nm Bi wires[3] have been interpreted in terms of transitions between L-band electron energy levels, which have been modified according to the predictions described in reference 1. Magnetoresistance results for $d > 6$ nm[4] indicate that the wires present evidence for quantum confinement. SdH studies of wires oriented along the C2 (bisectrix) direction, as described previously in reference 5, are very difficult because the electron periods are very long, and the various electron ellipsoids have different orientations with respect to the wire axis. For wires oriented along the trigonal direction, as was the case in this study, the SdH method is sensitive to short-period holes, which are easier to study within the experimental $1/B$ window. The hole SdH periods observed for 200-nm wires are longer overall than in the bulk, indicating a decreased $p$, a finding that is consistent with theoretical estimates of the effect of confinement on the FS.[6]

Bi and Bi-Sb crystals have the largest thermoelectric figure-of-merit of any material for $T <$ 100 K.[7] It has been predicted that Bi quantum wires can be engineered to possess enhanced thermoelectric properties because of the increase of the density of states resulting from quantum confinement along the wire axis.[8] Recent measurements of the thermopower, $\alpha$, of 9- to 15-nm Bi wires by Heremans *et al.* confirm this prediction.[9] However, it has been previously observed[9,10] that smaller-diameter wires (6 nm) do not show $\alpha$-enhancement, even though stronger effects are expected for the finest wires. Clearly, the confinement theory of small diameter Bi quantum wires requires some refinement.

Here we present an investigation of SdH oscillations of the longitudinal magnetoresistance (LMR) and transverse magnetoresistance (TMR) of 30-nm- and 80-nm-diameter Bi nanowires. Although 80-nm-diameter wires show an overall increase of the LMR and TMR SdH periods, consistent with confinement theory, the 30-nm wires do not follow the trend toward semiconducting behavior with increasing confinement. Instead, we observe that the SdH periods are much shorter, consistent with surface-induced carriers present in the nanowires. Additionally, the anisotropy of the hole Fermi surface measured is observed to decrease with decreasing wire diameter. Detailed angular studies of the 30-nm wires show that the SdH periods are essentially isotropic.

The arrays of Bi nanowires employed here are synthesized by injecting[11] molten Bi into anodic alumina templates. The hexagonal array of nanochannels in the templates is self-assembled by the anodization of Al. X-ray diffraction shows that more than 90% of the wires are oriented along the trigonal direction of the lattice structure of Bi. Figure 1a shows a scanning electron micrograph (SEM) of the 30-nm wires.

True four-probe resistance measurements are not feasible with nanowire arrays, and our two-probe technique consists of attaching Cu wire electrodes to both sides of the Bi nanowire composite by using silver epoxy contacts of about 1 mm$^2$ in size (Fig. 1c). We also utilized a quasi-four-probe technique,[6] in which the array is terminated in a thin (0.2 mm) layer of Bi. The experiments were performed on four samples. Samples *I* and *II* were 30-nm arrays, while sample *III* was an 80-nm array. Samples *I* through *III* were fitted with two-probe contacts. Sample *IV* was a 30-nm array in the quasi-four-probe configuration. Zero-field resistance and magnetoresistance measurements were made in two separate laboratories: (i) Boston College (Boston, Mass.) in a cryostat at a range of 1.8 < *T* < 300 K and *B* < 9 T and (ii) International Laboratory of High Magnetic Fields and Low Temperatures (Wroclaw, Poland) in a cryostat at a range of 1.8 < *T* < 300 K and *B* < 14 T, which allowed for the sample rotation around two Euler axes. Figure 1 shows the results for R(T) for two-probe samples for the three wire diameters used in this study. Typically, the resistance (R) of the two-probe samples was 1 Ω < R < 100 Ω. Due to the granular nature of the silver epoxy, electrical contact was made to a small fraction (~10$^{-5}$) of the wires in two-probe measurements. In comparison, the resistance of the 30-nm Bi-capped sample (*IV*), in which contact is made to all the wires in the array, was 600 μΩ at 1.8 K. The observed R(*T*) exhibits a negative temperature coefficient. The temperature dependence of the other samples was similar. The interpretation of the R(*T*) has been discussed elsewhere[4] and is problematic because separating changes in carrier density and mobility is difficult.

The Chambers magnetic field, $B_c$, for which the Landau level orbit diameter equals the wire diameter, is given by $B_c = \Phi_0 k_F/\pi d$, where $k_F$ is the Fermi wave vector and $\Phi_0 = hc/e$.[12] It is expected that LMR < TMR for $B > B_c(d)$ for wires in which electronic transport is ballistic because at high magnetic fields, the carriers avoid collision with the walls when the magnetic field is oriented along the wire direction. As is demonstrated with the LMR and TMR of sample *IV* (Fig. 1b), the LMR becomes increasingly smaller than the TMR for $B > B_c = 5$ T. Similarly, we find that $B_c \sim 2$ T for 80-nm wires. In comparison, we have found that $B_c \sim 1$ T for 200-nm wires. These values are in qualitative agreement with the wire diameter dependence of $B_c$, with a constant $k_F \sim 1.5 \times 10^6$ cm$^{-1}$, which contrasts with the decrease of $k_F$ for decreasing *d* as described in reference 4. SdH oscillations are apparent in Fig. 1b. SdH oscillations are periodic in an inverse magnetic field, and the period *P* is given by $e/(cA_F)$, where $A_F$ is the extremal cross-sectional area perpendicular to the magnetic field of any pocket of the FS. SdH oscillations appear for $B > \sim B_c$. The derivative of the TMR and LMR for our two-probe 80-nm Bi nanowires samples for $B > 0.3$ T versus $1/B$ is shown in Fig. 2. The maxima are indicated in the index plot in the inset. The LMR shows SdH oscillations with a period of $0.122 \pm 0.005$ T$^{-1}$. In comparison, the single-crystal bulk Bi hole LMR SdH periods for ***B*** // trigonal axis is 0.158 T$^{-1}$. The TMR period is $0.095 \pm 0.005$ T$^{-1}$. The accepted value of the period for hole carriers in bulk Bi for ***B*** ⊥ trigonal plane is 0.046 T$^{-1}$. LMR and TMR periods decrease in intensity with decreasing *B* near $B_c$, possibly due to carrier wall collisions.

Figure 3 shows the derivative of the LMR and TMR of two-probe 30-nm Bi nanowire samples *I* and *II* for $B > 2$ T. From the index plot (see inset, Fig. 3), we find that the periods of the LMR and the TMR of the samples are the same, with values of $0.025 \pm 0.002 \text{T}^{-1}$, which are much smaller than the SdH periods for holes in bulk Bi. Also, the SdH periods of holes and electrons in bulk Bi are highly anisotropic, a fact which contrasts with our observations here. We have also studied the SdH oscillation of sample *II* for various intermediate angles, i.e., the θ between the nanowires and the magnetic field using a two-axis rotator. The upper-left inset shows the angular variation of the SdH period, which is found to be essentially independent of the orientation of the nanowire with respect to the magnetic field. We also performed SdH measurements on sample *IV* (quasi-four-probe configuration) and found that the periods of the LMR and the TMR are the same, with a value of $0.030 \pm 0.002 \text{T}^{-1}$, indicating that the short period is not an artifact of the two-probe technique but is an intrinsic property of 30-nm-diameter Bi nanowires.

The data for the 80-nm Bi arrays in Fig. 2 can be interpreted in terms of the Bi FS model.[2] The number density for an ellipsoidal hole pocket[13] is $p = (8/3\pi^{1/2}) \Phi^{-3/2} (P_{h,//} P_{h,\perp}^2)^{-1/2}$. For bulk Bi, one finds that $p = 3.0 \times 10^{17}/\text{cm}^3$. We find for 80-nm Bi nanowires that $p = (8.6 \pm 0.3) \times 10^{16}/\text{cm}^3$, less than 30% of the hole density of bulk Bi. In comparison, our SdH study of 200-nm Bi-capped Bi nanowires[6] yielded $p = 2.6 \times 10^{17}/\text{cm}^3$. The substantial decrease of the spatial hole density for decreasing wire diameter indicates that the effect of confinement is to make the 200- and 80-nm wires approach the insulating state, the latter being closer to being a semiconductor. Assuming that the cyclotron mass is not changed and neglecting the change of the anisotropy, we find that $p \sim (E_o - E_F)^{3/2}$. Accordingly, we find that the observed decrease of the hole concentration can be considered to be caused by a 4.8-meV decrease of the overlap energy with respect to the Fermi energy, consistent with the predictions described in reference 1. However, our observations indicate that the ellipsoidal shape of the FS of Bi is modified by confinement to render it more spherical, which is contrary to the assumption in reference 1 of an effective mass tensor that is unchanged by such confinement.

30-nm Bi array samples show an isotropic SdH period of 0.025 T$^{-1}$, consistent with a spherical pocket of the Fermi surface of density $p = (1.3 \pm 0.2) \times 10^{18}$/cm$^3$. This finding is in sharp contrast with the prediction in reference 1 that the nanowires will be semiconducting for $d$ < 55 nm. The short period and the increased hole density are not particularly surprising in themselves. Increases of the hole density of Bi crystals can be engineered by adding electron acceptor impurities like Pb, resulting in shortened SdH periods.[13] It is plausible that the surface charges at the glass-Bi interface bring about a similar effect in nanowires. Since the Landau level diameter, which is presumably 30 nm for $B = B_c$, decreases with increasing magnetic field, field-independent SdH period indicates that the FS is the same throughout the interior of the nanowires. This finding agrees with the estimates of the screening length of a fixed charge in Bi of 30 nm.[14] While less important in the larger nanowires due to the decreased surface-to-volume ratio, surface states would also appear in the periphery of 80-nm wires, and a spatially nonuniform carrier density may explain the nonlinear index plot when $B^{-1} < 0.3$ T$^{-1}$. Thin Bi films have been well studied previously[15] and bear many similarities to Bi nanowires. The electronic transport measurements of thin Bi films have been interpreted in terms of a surface-induced hole charge ($p_S$) per unit area related to surface-induced states. The hole density is $p = p_i$ $+ p_S/t$, where $p_S = 3 \times 10^{12}/cm^2$ and $t$ is the film thickness for very thin films ($t < 100$ nm). Considering that it is proportional to the specific surface area, we estimate that $p = 2 \times 10^{18}/cm^3$ for $d = 30$ nm. Therefore, the observation of surface-induced hole density in thin films is consistent with the hole density data presented here for 30-nm wires.

The value for $m^*$ can be estimated from the temperature dependence of the SdH amplitude[4] and is found to be 0.3 $m_e$, while in comparison, the average effective mass of holes for bulk Bi is 0.065 $m_e$. Since, neglecting spin, the SdH period of the heavy carriers is given by $\mu_B/[m^*(E_o - E_F)]$, we find that $(E_o - E_F) = 8$ meV, a value which is roughly the same as that for the 200-nm wires. Therefore, the reason for the small period of the heavy carriers is the large increase of the effective mass from the value for bulk Bi.

The trend toward decreased FS anisotropy for decreasing wire diameter is more puzzling than the high value of $p$ in 30-nm wires. The FS anisotropy is typically unaffected by changes in hole density resulting from the addition of acceptor and donor impurities in bulk Bi.[2,13] It is possible that the different symmetry of the spatial variation of the potential of the carriers, which is central symmetric in the single crystal and radial in nanowires, can account for this difference.

Experiments similar to the one presented here for 30-nm Bi wires were performed in 6-nm and 18-nm Bi wires. These samples do not exhibit SdH oscillations. This is not surprising, given that the smaller wires will exhibit a reduced MR and higher effective masses.

In summary, the SdH measurements on arrays of 80-nm wires show a significantly reduced hole carrier density that is consistent with a confinement model[1] and the approach to a SMSC transition as the wire diameter decreases. However, the isotropy of the FS increases with decreasing wire diameter. Also, 30-nm-diameter wires are not semiconductors as predicted. Instead, we observed a FS pocket that is nearly spherical, with a large charge density and relatively large effective mass. The increased carrier density is believed to be surface-induced.

The 80-nm template was kindly provided by H. Masuda of Tokyo Metropolitan University. This work was supported by the National Science Foundation, the Army Research Office, and the Civilian Research Development Fund.

±     Current Address: Trex Hawaii LLC, 3038 Aukele Street, Lihue, HI 96766

**FIGURE CAPTIONS**

Fig. 1.     Temperature dependence of the resistance of the two-probe 30-nm, 80-nm, and 200-nm wire arrays. The various samples employed are indicated: a) SEM side-view of the 30-nm Bi wire array, b) MR of sample *IV*, and c) the 2 K transverse and 2 K, 5 K, 10 K, and 20 K longitudinal magnetoresistance MR = R(*B*) – R(0) of the 30-nm Bi nanowires as indicated; $B > 0$ curves have been shifted vertically.

Fig. 2.     The LMR and TMR of the 80-nm Bi nanowires at 1.8 K (solid line) and 5 K (dashed line). Inset: oscillation index versus 1/*B* for sample *III*.

Fig. 3.     Angle-dependent derivative of the MR of 30-nm Bi wire array samples as indicated. Lower inset: *n* versus 1/*B* for sample *I*; triangles and circles indicate θ = 90º and 0º, respectively. Upper inset: polar plot of the SdH periods of sample *II* with the gray arc corresponding to 0.025 $T^{-1}$.

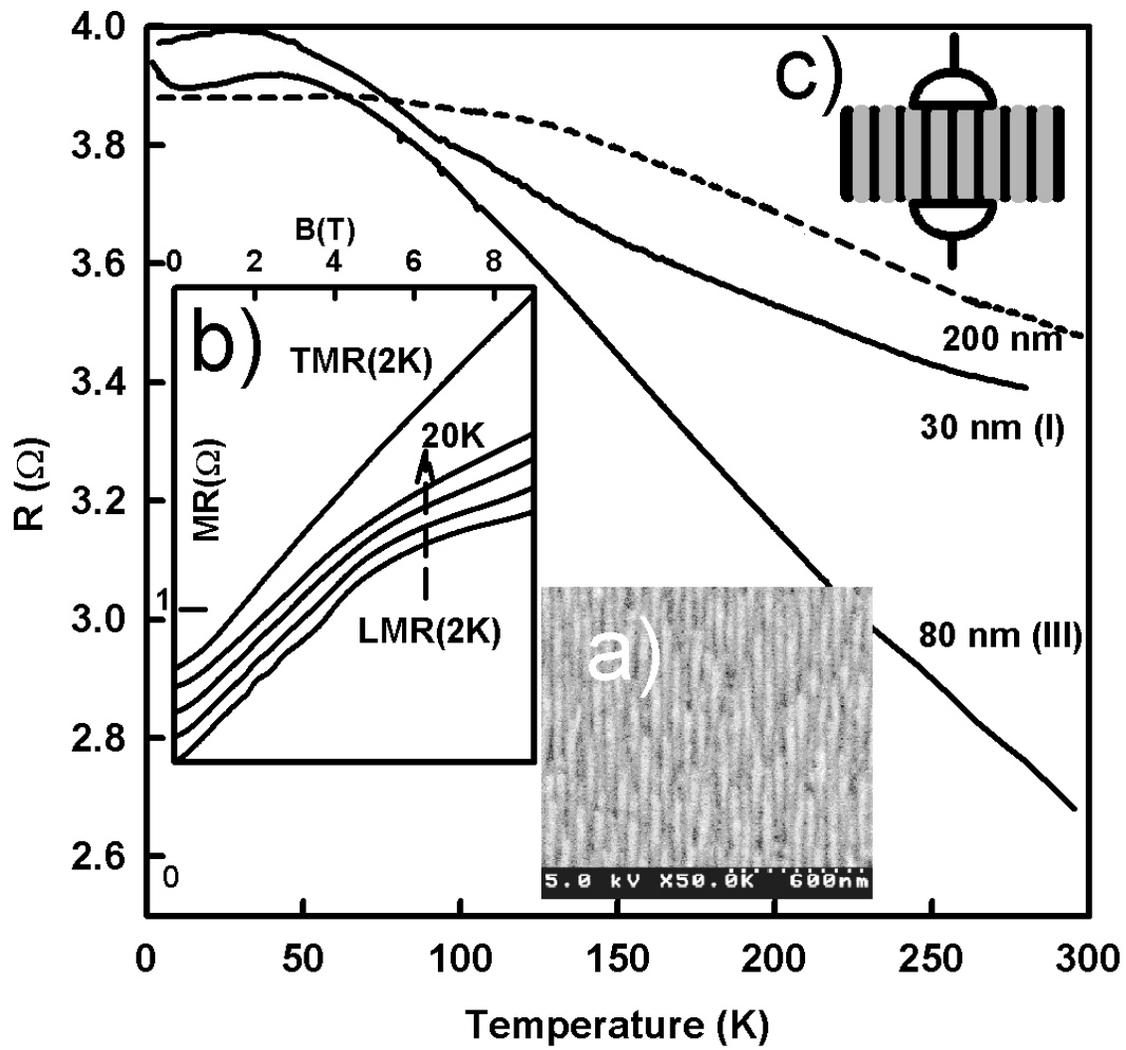

Fig. 1. Huber. APL

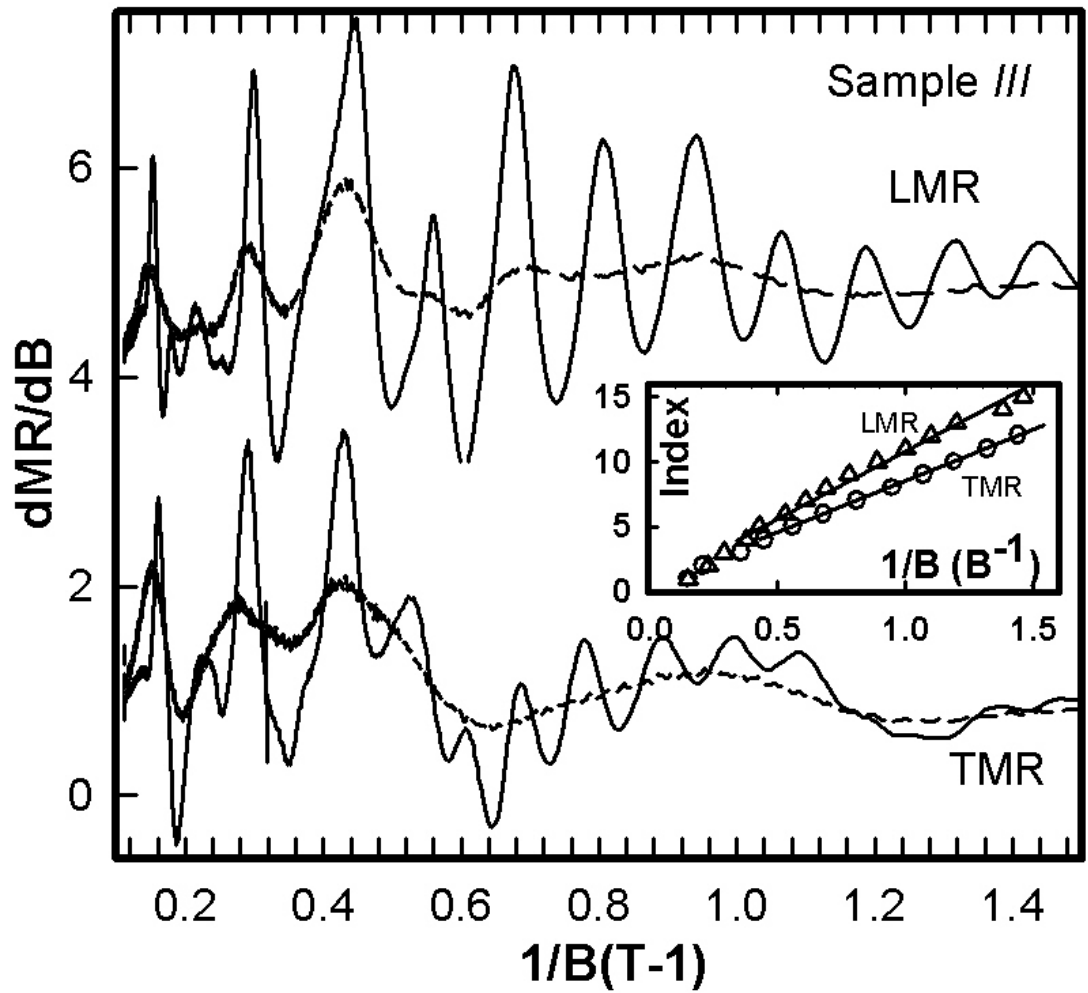

Fig. 2. Huber. APL

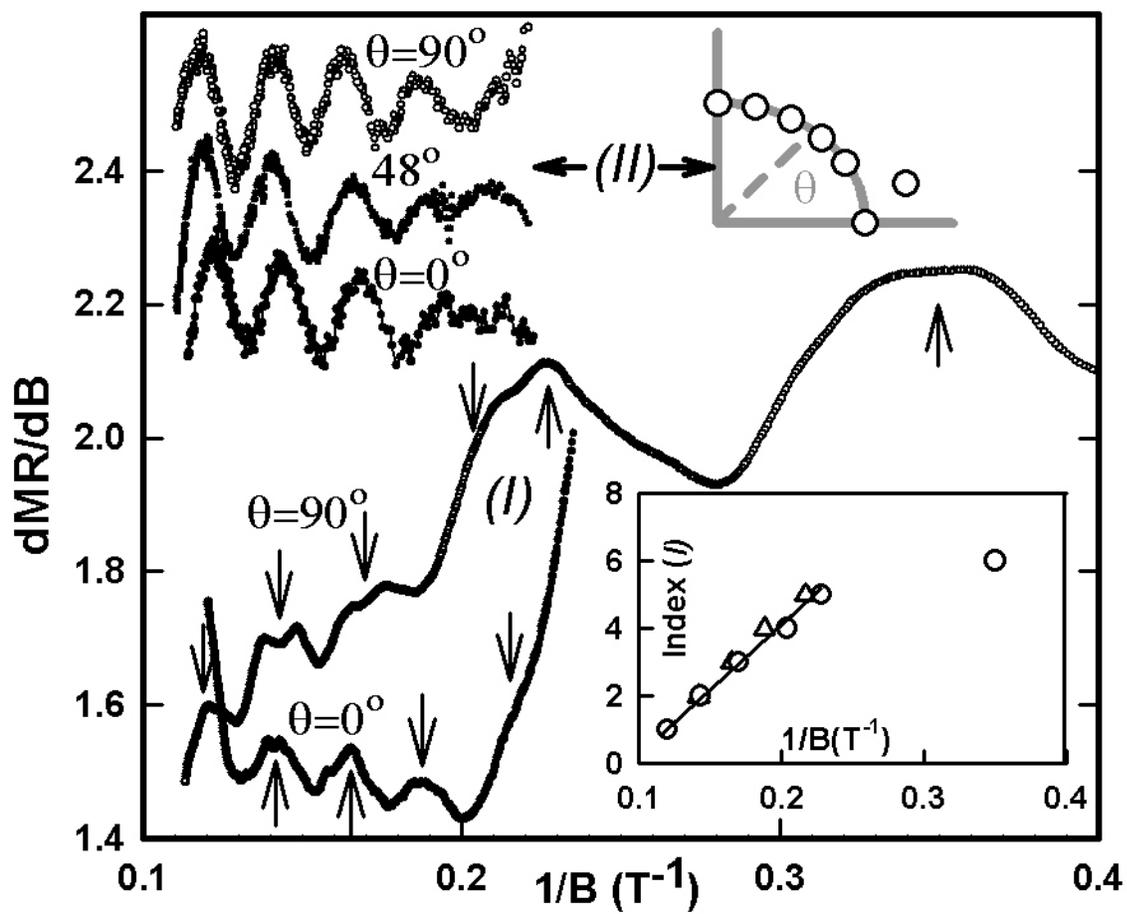

Fig. 3. Huber. APL